\documentclass{caosp}
\usepackage{graphicx}

\articleNo{123}
\pubyear{2008}
\volume{38}
\volnumber{1}
\firstpage{1}
\received{October 30, 2007}
\accepted{November 30, 2007}

\begin{document}

\hauthor{A.\,Sapar, A.\,Aret, L.\,Sapar and R.\,Poolam\"{a}e}

\title{Formulae for study of LID induced diffusion \\in CP star model atmospheres}

\author{
        A.\,Sapar
      \and
        A.\,Aret
      \and
        L.\,Sapar
      \and
        R.\,Poolam\"{a}e
       }

\institute{
           Tartu Observatory\\
           61602 T{\~oravere}, Estonia \email{sapar@aai.ee}
          }

\date{October 30, 2007}

\maketitle

\begin{abstract}
Formulae suitable for computing LID acceleration and corresponding  diffusional
segregation of isotopes in CP star model atmospheres are given.
\end{abstract}

\noindent Light-induced drift (LID) as a phenomenon responsible for
separation of  isotopes in the atmospheres of CP stars has been
proposed by Atutov and Shalagin (1988). Thereafter we have studied
evolutionary abundance changes of Hg and its isotopes (see Sapar
{\it et al.}, C\_Sapar, these proceedings). Here we give formulae describing
diffusion of isotopes in the form best suited for numerical model
computations.

Diffusive transfer of radiation holds in the deeper layers of stellar atmospheres,
 where monochromatic radiative flux can be described by formula
{\abovedisplayskip=4pt \belowdisplayskip=4pt
\begin{displaymath}
  F_\nu= {F K_R}\frac{dB_\nu}{\kappa_\nu dT}~,~~~~~~
 {\rm where}~~ \frac{1}{K_R}= \int_0^\infty \! \frac{dB_\nu}{\kappa_\nu dT} d\nu.
\end{displaymath}
}Here $F$ is the total radiation flux and $K_R$ is the Rosseland opacity integral.
In the opacity coefficient $\kappa_\nu$  the overlapping  spectral
lines of the trace element studied play an important role, and thus
$
     \kappa_\nu = c_\nu+\sum_j  \sigma_j W_j(u_\nu),
$
where $c_\nu$ is the continuous opacity coefficient and $\sigma_j$ is the transition cross-section per gram.
Summation is made over spectral lines $j$ with line profile functions $W_j$, usually
being the Voigt functions with argument given by
  $u_{\nu }=(\nu -\nu _j)/\Delta \nu _{T}$,
where $\Delta \nu _{T}$ is the thermal Doppler width of the spectral line.
 As we have shown, the effective acceleration of LID due to spectral line $j$ can be expressed
quite similarly to the usual expression of radiative acceleration
{\abovedisplayskip=4pt \belowdisplayskip=4pt
\begin{equation}
a_j^L=\frac{\pi\varsigma_j}{c} \int_0^\infty \frac{\partial
W_j(u_\nu)}{\partial u_\nu}F_\nu d\nu, ~~~~~~~\varsigma_j=q\epsilon
\sigma_j,
\label{eq:acLID}
\end{equation}
}where $q=Mv_Tc/2 h\nu$ and the efficiency of LID is
$\epsilon=(C_u-C_l)/( A_u+C_u)$. Here $C_u$ and $C_l$ are the
collision rates of particles in upper and lower states,
respectively, and $A_u$ is the probability of spontaneous
transitions.

Expression (\ref{eq:acLID}) in the region of diffusive transfer of radiation takes the form
{\abovedisplayskip=4pt \belowdisplayskip=4pt
\begin{equation}
a_j^L=\frac{\pi\varsigma_j F K_R}{c
}\int_0^\infty\frac{\partial W_j(u_\nu)}{\partial u_\nu}
{dB_\nu\over \kappa_\nu dT} d\nu.
\label{eq:aLID}
\end{equation}
}Both isotopic and hyperfine splitting of spectral lines of all ions should be taken into account
to calculate accelerations producing segregation of isotopes.

The equation of  continuity for isotope $i$ in the plane-parallel
stellar atmosphere has the form $ {d\rho_i}/{dt}+ {d(\rho_i
V_i)}/{dr}=0. $ The model atmosphere data correspond to standard
points, being equidistant on the logarithmic scale of mean optical
depth. These points are enumerated as layers $n$ growing downwards.
Treating $n$ as  a continuous parameter, we change variables in
equation of continuity. Since
 $ \frac{d}{dr}=\frac{d/ dn}{dr/ dn} $ and
$\rho\frac{ dr}{dn}=-\frac{d\mu}{dn}$,
 where $\mu$ is total column
density, we get for radial  gradient {\abovedisplayskip=4pt
\belowdisplayskip=4pt
\begin{displaymath}
\frac{d}{dr}= -\gamma \frac{d}{dn}, ~~~{\rm where}~ \gamma = \frac{\rho}{d\mu/dn} = \frac{\rho}{\mu d\ln \mu/dn}~.
\end{displaymath}
}Denoting a ratio of current concentration to its initial value as $C_i$,  we can write $\rho_i=\rho_i^0 C_i$ and
 the equation of continuity reduces to
{\abovedisplayskip=4pt \belowdisplayskip=4pt
\begin{equation}
\frac{d\ln C_i}{dt} = \frac{\gamma}{\rho_i}\frac{d(\rho_i V_i)}{dn}~.
\label{eq:dlndc}
\end{equation}
}Logarithms are used to avoid possible negative values of $C_i$ in time integration.
Diffusion velocity $V_i$ in the presence of stellar wind can be found from equation
{\abovedisplayskip=4pt \belowdisplayskip=4pt
\begin{equation}
\rho_i V_i=\rho_i (a_i-g)t  -\Delta\frac{d\rho_i}{dr}~,
~~~~~\frac{d\rho_i}{dr} = -\gamma\frac{d\rho_i}{dn},
 \label{eq:rhov}
\end{equation}
}where $a_i$ is the sum of radiative and LID accelerations, $t$ is
the mean free flight time of the particles~, $m$ is the mean mass of
buffer particles and $\Delta=kTt/m$ is the diffusion coefficient of
trace particles. Thus from equation~(\ref{eq:rhov})  we find
{\abovedisplayskip=4pt \belowdisplayskip=4pt
\begin{equation}
\frac{V_i}{\gamma\Delta} = \frac{m(a_i-g)}{k T\gamma} + \frac{d\ln
(\rho_i^0C_i)}{dn}~~~~{\rm and}~~~~
\frac{V_i}{\gamma\Delta}  = \frac{ma_i}{k T\gamma} +  \frac{d\ln C_i}{dn}.
 \label{eq:dlndn}
\end{equation}
}Here the last expression has been obtained from the first
taking into account that for model stellar atmospheres approximately holds
$mg / kT\gamma = d\ln \rho_i^0 / dn$.
The velocity $V_i$  is to be used in the equation of continuity (\ref{eq:dlndc}), reducing it to a
generalized Fokker--Planck equation. The  equation (\ref{eq:dlndn}) can be
used also for a crude prediction  of concentrations $ C_i$. For the final
equilibrium state in the case of no stellar wind $V_i=0$ and in the
case of constant mass loss rate $\rho_i V_i = \dot m_i$.

 As we see, for evolutionary
computations we need to find  derivatives $d\ln\rho / dn$,
 $d\mu /
dn$ and  $d\gamma / dn$,
  which correspond to buffer gases and several derivatives for each isotope of the
trace (impurity) particles.
The derivatives have been
  found using the 4th order Lagrange interpolation formulae for equidistant nodes.

\acknowledgements
We are grateful to Estonian Science Foundation for financial support by grant ETF 6105.



\begin{thebibliography}{}
\article{Atutov, S.N., Shalagin, A.M.}{1988}{\sal}{14}{284}
\bibitem{}C\_Sapar, these proceedings

\end{thebibliography}
\end{document}